# Comparison between security majors in virtual machine and linux containers

Udit Gupta
Information Networking Institute
Carnegie Mellon University, Pittsburgh – Pennsylvania, USA
uditg@andrew.cmu.edu

**Abstract:**

Virtualization started to gain traction in the domain of information technology in the early 2000's when managing resource distribution was becoming an uphill task for developers. As a result, tools like VMWare, Hyper-V (hypervisor) started making inroads into the software repository on different operating systems. VMWare and Hyper-V could support multiple virtual machines running on them with each having their own isolated environment. Due to this isolation, the security aspects of virtual machines (VMs) did not differ much from that of physical machines (having a dedicated operating system on hardware). The advancement made in the domain of linux containers (LXC) has taken virtualization to an altogether different level where resource utilization by various applications has been further optimized. But the container security has assumed primary importance amongst the researchers today and this paper is inclined towards providing a brief overview about comparisons between security of container and VMs.

## 1. Introduction:

The concept of virtualization [1 and 2] was created with the sole intent of managing resource distribution efficiently and hence system resources started getting divided logically instead of physical division. Ever since the introduction of mainframe computers in 1960s by IBM, this technique of logical division has been of great utility. Apart from efficient resource distribution, virtualization has another advantage: it is easier to deal with systems at software level than working on hardware. Hardware virtualization has enabled users to deal with hardware more efficiently via logical connections where a hypervisor emulates a piece of computer hardware.

With the advent of cloud computing [3 and 4], virtualization has assumed primary importance for enterprises as well. It becomes easier to manage data centers across the world and thus it helps in working remotely. As functionality tends to increase for enterprise applications, there has been a disproportionate rise in the number of virtual machines (VMs) on each data center. Each hypervisor had an upper limit on the number of VMs that can run on it. Moreover, most of these applications did not require even half the resources allocated to VM by CPU. Thus, it was mandated with growing number of applications to run two to three times more server instances on a given server as compared to VMs. This lead to the development of linux containers (LXCs) [15 and 16] wherein only those resources will be used which are required by applications.

Although LXCs has been used efficiently in successfully executing various enterprise level applications, the security features of LXCs are yet to evolve when compared with VMs. Due to the isolation provided in VMs the threat of intrusions has been neutralized to a great extent via intrusion detection algorithms [5, 6, 7 and

8]. Furthermore, all the security aspects pertaining to detection of spam [9] and malware [10 and 11] which have been addressed in VMs are yet to remain addressed in containers. Although each LXC receives its own network stack and process space as well as its instance of file system, it doesn't have its own user namespace. As we proceed further we'll look at various container based technologies how security features varies across LXCs and VMs.

## 2. Container based technology:

Linux containers (LXCs) provide operating system-level virtualization and they have their own process and network space. Although work is still going on in creating a separate user namespace, the isolation provided in other aspects has been at par with virtual machines. Keeping in view the demand of growing number of instances required along with sandboxing, various container based technologies were designed to address this feature. The two primary container based technologies which we'll look at: docker and openVZ.

**2.1 OpenVZ:** OpenVZ [17 and 18] is a container based technology which allows a physical server to run multiple instances of an operating system called containers or virtual private servers (VPS). Each container has its own network stack, serial ports, process tree and file system. In later versions of openVZ, work is being done to create container's own user space. It uses single patched linux kernel and can run only linux. It might be disadvantageous to use openVZ in case containers need different kernel version since all openVZ containers share the same architecture and kernel version. Also since it doesn't carry the overhead associated with operating system, it is much more efficient, scalable. Furthermore, there is dynamic memory allocation i.e. the memory allocated to 1 linux container can be used for other container as well without the requirement of rebooting the entire system.

**2.2 Docker:** Docker [14] is different from openVZ in the sense that former sees a container as an application/service while latter sees container as a VPS. Since container is a single application in docker's terminology, it is important that interface between various containers needs to be robust since multiple containers might be required to run an application. For instance, in order to run a particular application, we might need configuration files from 1 container and database from another container. Thus, it is important to have secure and robust communication between two containers. Since each container has its own network stack, the secure communication between 2 containers can be achieved via any of the following protocols: SFTP [19], SSH [20], FTPS [21 and 22] and SCP.

Docker also has the capability of importing/exporting containers via access to the public registry. Moreover, docker defines an API for automating and customizing the creation and deployment of containers. It also has the capability of tracking and managing successive versions of a container, inspecting diff between versions, committing new versions, etc. Apart from this it has similar features compared to openVZ where it provides namespace, file system, network, resource isolation. Furthermore, docker also provides logging feature where the standard streams of each process container is collected and logged for real-time or batch retrieval. Lastly, docker provides an abstraction layer for containers so that they can run on different operating systems without any compatibility issues.

## 3. Comparison between containers based technology and virtual machines

Linux containers were designed with a single view of managing CPU resources distribution more efficiently. On any instance of VMWare [12] or Hyper-V, it is difficult to run more than 10 VMs due to the overhead incurred. Containers have resolved this problem to a great extent where they only make use of resources which are required by the application or service. Thus, more than 50 instances of container can run on single quad-core machine. Let's consider the example of any enterprise email security product: its main functionality would be to scan emails for spam/virus/malware, manage logs [13], manage message transfer agent (MTA) and report any datacenter outage in case the product is deployed on cloud network. In most cases, these functionalities described will not make any use of kernel data structures or operating system libraries or any related dependencies. Thus, rather than having VMs for each aspect of the product, it's better to containerize each feature by sandboxing them using docker/openVZ. In many of the organizations, VMs are emulated so as to perform feature testing which consumes hordes of memory space and CPU utilization. Having containers would ensure that redundancy would not have much impact on the resource consumption. Scalability would be easier since the time required for container installation is much less as compared to VMs.

On the other hand security aspect of docker/openVZ has been of concern lately. As isolation reduces, security is bound to decrease exponentially. Since containers share the same operating system and kernel, it is easier to gain access to containers especially as root user on linux. Although docker isolates many aspects of the underlying host from an application running in a container from an underlying host, the separation is not as strong as that of VM. Moreover, some applications might need to run on different operating system which is not possible in containerized technology.

Keeping in view the advantages and disadvantages of LXCs/docker, we need to arrive at some kind of trade off. The ideal situation would be to have few VMs installed on physical machine and then have many instances of container running on VM. This would ensure that security features of VM along with optimized features of LXCs would give maximum performance for a system. In case of cloud networks having data centers around the world, few VMs can be installed on a base machine on every datacenter followed by running multiple instances of LXCs on those VMs. Since resource consumption would reduce drastically, this solution would be cost effective for many organizations.

## 4. Conclusion:

The manner in which virtualization has dominated over the past decade goes on to show how important the role of linux based containers will be in future. This is evident from the fact that many organizations like Amazon, Microsoft have already adopted this technology in their cloud based products. Furthermore, the scalability which is being offered by the container technologies would ensure that cost savings would grow manifold in the near future. In this paper, a brief overview was provided about docker, openVZ: two container based technologies which are often seen as potential replacement for VM. In addition, a comparison was provided between LXCs and VMs from general as well security point of view.